\documentclass[conference, 10pt]{IEEEtran}
\IEEEoverridecommandlockouts
\usepackage{cite}
\usepackage{amsmath,amssymb,amsfonts}
\usepackage{mathtools}
\usepackage{algorithm}
\usepackage{algorithmic}
\usepackage{graphicx}
\usepackage{textcomp}
\usepackage{xcolor}
\usepackage{hyperref}
\usepackage{pgfplots, tikz}
\usepgfplotslibrary{groupplots}
\usepackage{enumitem}
\usepackage{caption, subcaption}
\pgfplotsset{compat=1.12}

\usepackage[left=0.65in,right=0.65in,top=0.75in,bottom=1.05in]{geometry}
\usepackage{amsthm, bm, siunitx} 
\DeclareSIUnit{\belmilliwatt}{Bm}
\DeclareSIUnit{\dBm}{\deci\belmilliwatt}

\DeclareMathOperator*{\trace}{tr}
\newcommand{\tr}[1]{\trace\left(#1\right)}

\newcommand{\normsq}[1]{\left\lVert#1\right\Vert_{2}^{2}}

\newcommand{\cov}[1]{\bm{C}_{#1}}

\newcommand{\gauss}[1]{\mathcal{N}\left(\bm{0}, #1\right)}

\newcommand{\eye}{\bm{\mathrm{I}}}

\newcommand{\herm}{^{\mathsf{H}}}

\newcommand{\C}{\mathbb{C}}

\newcommand{\x}{\bm{x}}
\newcommand{\y}{\bm{y}}
\newcommand{\ch}{\bm{H}}

\newcommand{\cg}{\bm{G}}
\newcommand{\z}{\bm{z}}

\newcommand{\wa}{\bm{w}}

\newcommand{\shb}{\hat{s}}

\newcommand{\vb}{\bm{v}}

\newcommand{\bmeta}{\bm{\eta}}

\newcommand{\btheta}{\bm{\theta}}
\newcommand{\bpsi}{\bm{\psi}}

\newcommand{\bmb}{\bm{b}}

\newcommand{\bu}{\bm{u}}

\newcommand{\nt}{N_{T}}
\newcommand{\ntq}{N_{T_{q}}}
\newcommand{\nr}{N_{R}}

\newcommand{\nsc}{N}

\newcommand{\symset}{\mathcal{K}}
\newcommand{\carrierset}{\mathcal{N}}
\newcommand{\resourceset}{\mathcal{R}}
\newcommand{\userset}{\mathcal{Q}}

\def\BibTeX{{\rm B\kern-.05em{\sc i\kern-.025em b}\kern-.08em
    T\kern-.1667em\lower.7ex\hbox{E}\kern-.125emX}}
    
\begin{document}

\title{

Resilient-By-Design Framework for MIMO-OFDM Communications under Smart Jamming

\thanks{
The authors were supported in part by the
German Federal Ministry of Education and Research (BMBF)
in the programme “Souver\"an. Digital. Vernetzt.”
within the research hub 6G-life under Grant 16KISK002,
and also by the Bavarian Ministry of Economic Affairs,
Regional Development and Energy within the project 6G Future Lab Bavaria.
U. M\"onich and H. Boche were also supported by the BMBF within the project "Post Shannon Communication - NewCom" under Grant 16KIS1003K.
W. Saad was supported by the Center for Assured and Resilient Navigation in Advanced TransportatION Systems (CARNATIONS) under the US Department of Transportation (USDOT)'s University Transportation Center (UTC) program (Grant No. 69A3552348324)."
}
}

\author{\IEEEauthorblockN{
Vlad C. Andrei\IEEEauthorrefmark{1},
Aladin Djuhera\IEEEauthorrefmark{1},
Xinyang Li\IEEEauthorrefmark{1}, 
Ullrich J. M\"onich\IEEEauthorrefmark{1},
Holger Boche\IEEEauthorrefmark{1}
and Walid Saad\IEEEauthorrefmark{2}
\IEEEauthorblockA{\IEEEauthorrefmark{1}Chair of Theoretical Information Technology, Technical University of Munich, Munich, Germany\\
\IEEEauthorrefmark{2}Electrical and Computer Engineering Department, Virginia Tech, Arlington, USA, \\
Emails: \{vlad.andrei, 
aladin.djuhera,
xinyang.li,
moenich,
boche\}@tum.de,  
walids@vt.edu
}  }}

\maketitle

\begin{abstract}
Native jamming mitigation is essential for addressing security and resilience in future 6G wireless networks. In this paper a resilient-by-design framework for effective anti-jamming in MIMO-OFDM wireless communications is introduced. A novel approach that integrates information from wireless sensing services to develop anti-jamming strategies, which do not rely on any prior information or assumptions on the adversary's concrete setup, is explored. To this end, a method that replaces conventional approaches to noise covariance estimation in anti-jamming with a surrogate covariance model is proposed, which instead incorporates sensing information on the jamming signal's directions-of-arrival (DoAs) to provide an effective approximation of the true jamming strategy. The study further focuses on integrating this novel, sensing-assisted approach into the joint optimization of beamforming, user scheduling and power allocation for a multi-user MIMO-OFDM uplink setting. Despite the NP-hard nature of this optimization problem, it can be effectively solved using an iterative water-filling approach. In order to assess the effectiveness of the proposed sensing-assisted jamming mitigation, the corresponding worst-case jamming strategy is investigated, which aims to minimize the total user sum-rate. Experimental simulations eventually affirm the robustness of our approach against both worst-case and barrage jamming, demonstrating its potential to address a wide range of jamming scenarios. Since such an integration of sensing-assisted information is directly implemented on the physical layer, resilience is incorporated preemptively by-design.
\end{abstract}


\section{Introduction}
The advent of 6G networks is set to advance wireless communications with the introduction of novel applications such as immersive extended reality, distributed learning and ultra-precise localization services \cite{schwenteck_2023_6g}. In order to guarantee security and ensure user trust, such services will inevitably demand more stringent and increasingly challenging requirements regarding resilience and trustworthiness of the wireless system design \cite{fettweis_boche_6g_bits}. A critical aspect of ensuring this trust is the implementation of robust protection schemes against adversarial jammers, who may seek to severely disrupt these services. In particular, resilient-by-design systems \cite{fettweis_boche_resilience2022} will be needed, i.e. systems, which preemptively provide protection against denial-of-service (DoS) attacks, as well as inherently detect, remediate and recover from those. 
Furthermore, with the continuous fusion of data and services in 6G, wireless sensing services may offer an additional wealth of valuable information.
This motivates the study of whether such sensing information can be harnessed to augment existing jamming mitigation schemes \cite{boche_schaefer_comp_feedback}.

Prior works in the context of anti-jamming, while substantial, have traditionally either only studied mitigation techniques for OFDM and multi-user MIMO systems separately, as in \cite{pirayesh_jamming_2021, shahriar_phy-layer_2015, do_jamming-resistant_2018}, or imposed strong assumptions on the adversary's knowledge and setup. For example, recent work in \cite{marti_mitigating_2023} and \cite{marti_universal_2023} only consider single- or few-antenna jammers with common secrets being exchanged between legitimate parties. This essentially excludes worst-case jammers, i.e. the most disruptive kind with comprehensive system knowledge, as studied in \cite{disrupting_mimo_comms, worst_case_jamming_mimo, corr_jamming, mimo_jammer_mac_bc}. Thus, we move away from such strong and potentially unrealistic assumptions towards intelligent and reconfigurable worst-case jammers with arbitrary antenna configurations and develop a universally applicable anti-jamming methodology on the basis of sensing-assisted information on the jamming signal DoAs. Additionally, we extend our studies by further considering optimal scheduling and resource allocation, filling the gap in the literature regarding the integration of anti-jamming strategies with the complexities of joint beamforming, scheduling, and resource allocation in multi-user MIMO-OFDM channels \cite{shahriar_phy-layer_2015}. 

In the following sections, a sensing-assisted anti-jamming framework is developed solely on the basis of available DoA information at the legitimate transceivers, whereno limitations are imposed onto the jammer.
To this end, first the wireless system model for multi-user MIMO-OFDM channels is introduced. Second, jamming resilience is incorporated into the system by means of an optimization objective, which maximizes the user sum-rate at the legitimate parties. In this problem, any explicit knowledge about the jamming strategy is replaced by a surrogate expression, which is dependent on the jamming signal DoAs only. The optimization problem is then solved using an iterative water-filling approach \cite{yu2004iterative}. In order to verify the effectiveness of the solution, the worst-case jamming strategy is developed with full system knowledge at the adversary and the anti-jamming framework is benchmarked against it, yielding substantially better performance. This introduces a novel, sensing-assisted approach to anti-jamming, which is robust even against the worst-case of adversarial jamming. 

\section{System Model and Problem Statement}\label{sec:sys_model}
We consider a MIMO-OFDM uplink scenario involving $Q$ legitimate users (UEs), one legitimate base station (BS), and one jammer. In every slot, each user $q \in \mathcal{Q} = \{1, \hdots, Q\}$ is allocated a resource set $\mathcal{R}_q = \mathcal{N}_q \times \mathcal{K}_q$ consisting of a subcarrier set $\carrierset_q$ and a symbol set $\symset_q$, which satisfy $\cup_{q=1}^{Q} \carrierset_{q} \subseteq \carrierset = \{1, \dots, \nsc\}$ and  
$\cup_{q=1}^{Q} \symset_{q} \subseteq \symset = \{1, \dots, K\}$, where $\nsc$ and $K$ denote the total number of subcarriers and OFDM symbols available.
Each user $q$ subsequently maps his zero-mean, unit-variance Gaussian data symbol $s_{qnk}$ onto the resource element (RE) and transmits the signal $\x_{qnk} = \alpha_{qnk}\sqrt{p_{qnk}}\wa_{qnk} s_{qnk}$ through his $\ntq$ antennas. Here $\wa_{qnk}$, $p_{qnk}$ denote the beamforming vector and the power allocation, while $\alpha_{qnk} \in \{0,1\}$ indicates whether data is scheduled at subcarrier $n$ and OFDM symbol $k$. Without loss of generality we assume $\ntq = \nt, \,\forall q\in \mathcal{Q}$.

The legitimate transmit signals propagate through the channels $\ch_{qnk}$ before arriving at the BS, which is equipped with $\nr$ receive antennas. The received signal is further corrupted by additive white noise $\bmeta_{nk}\sim\gauss{\sigma^2\eye}$ and the worst-case jamming signal $\bu_{nk} \sim \gauss{\cov{\bu_{nk}}} \in \C^{}$, which propagates through a separate jamming channel $\cg_{nk}$. More formally,
\begin{align}
    &\z_{nk} = \cg_{nk}\bu_{nk} + \bmeta_{nk},\\
    &\y_{nk} = \sum_{q\in\mathcal{Q}}\ch_{qnk}\x_{qnk} + \z_{nk} \in \C^{\nr}.
\end{align}
Finally, the BS forms estimates $\shb_{qnk} = \vb_{qnk}\herm\y_{nk}$ of the data symbols on each RE by applying the linear equalizer $\vb_{qnk}\herm \in \C^{1 \times \nr}$. Both $\ch_{qnk}$ and $\cg_{nk}$ are modeled as beamspace channels, that is
\begin{align}\label{eq:beamspace1}
    &\ch_{qnk} = \sum_{l=1}^{L_{H_{q}}} 
    b_{H_{q}, lnk} 
    \bm{a}_{\nr}(\btheta_{q,l}) \bm{a}_{\ntq}\herm(\bpsi_{q, l}), \\\label{eq:beamspace2}
    &\cg_{nk} = \sum_{l=1}^{L_{G}} 
    b_{G, lnk}
    (\btheta_{G, l}) \bm{a}_{N_{J}}\herm(\bpsi_{G,l}).
\end{align}
Here, $L_{H_{q}}$, $L_{G}$ indicate the number of resolvable paths, $\bm{a}_{\cdot}(\btheta)$ the steering vectors, and $\btheta_{\cdot, l}$, $\bpsi_{\cdot, l}$, $\alpha_{\cdot, l}$ the azimuth and elevation directions-of-arrival, directions-of-departure, respectively. The terms $b_{H_{q}, lnk}$ and $b_{G, lnk}$ include the contribution of the path gains, propagation delays and Doppler shift of each multipath component $l$. 

We assume the legitimate parties have perfect knowledge of their link parameters, 
i.e. $\{\alpha_{qnk}, p_{qnk}, \wa_{qnk}, \vb_{qnk}, \ch_{qnk}, \sigma^2\}$ for all $q \in \userset$ and $(n, k)\in\resourceset_q$, and of the jamming signal DoAs $\bm{\theta}_{G} = \{\theta_{G, l}\}_{l=1}^{L_{G}}$, which might be estimated by the BS and fed back to the UEs using bistatic sensing services in 6G \cite{schwenteck_2023_6g}.
For the receiver we consider the \textit{user-sum-rate} metric, defined as
\begin{align}
    &R^{B} = \sum_{\substack{(n,k)\in\resourceset \\ q\in\userset}} \alpha_{qnk} I\left(\shb_{qnk}; s_{qnk}\right) \\
    \label{eq:rate_bob}
    &=\sum_{\substack{(n,k)\in\resourceset \\ q\in\userset}}
    \log\left( 1 + \dfrac{\lvert\vb_{qnk}\herm\ch_{qnk}\bmb_{qnk}\rvert^2}{\vb_{qnk}\herm\bm{X}_{qnk}\vb_{qnk}}
    \right),
\end{align} 
where $I(\x;\y)$ is the mutual information between two random variables and we defined the interference-plus-noise covariance matrix
\begin{align}
    &\bm{C}_{\bm{z}_{nk}} = \cg_{nk} \cov{\bu_{nk}} \herm{\cg_{nk}} + \sigma^2 \bm{I}_{N_R} \label{eq:noise_cov}\\
    \label{eq:interf_plus_noise}
    &\bm{X}_{qnk} = \sum_{q^{\prime}\neq q} \ch_{q^{\prime}nk}\bmb_{q^{\prime}nk}\bmb_{q^{\prime}nk}\herm\ch_{q^{\prime}nk}\herm + \cov{\z_{nk}},
\end{align}
with $\bmb_{qnk} = \sqrt{p_{qnk}}\alpha_{qnk}\wa_{qnk}$. 
The unnormalized optimal equalizers can be then found in closed form as 
\begin{align}
    \label{eq:user_rx}
    \vb_{qnk}^{B} &= \bm{X}_{qnk}^{-1}\ch_{qnk}\bmb_{qnk}
\end{align}
The transmit parameters are chosen such that they maximize the \textit{sum-rate} $R^{A}$:
\begin{align}
    &R^{A} = \sum_{(n,k) \in \resourceset} I\left( \y_{nk}; \{s_{qnk}\}_{q\in\userset}\right) \\ 
    &=\sum_{(n,k) \in \resourceset}\log\left(
        1 + \sum_{q\in\userset} \alpha_{qnk} p_{qnk}\gamma^{A}_{qnk}(\cov{\z_{nk}})
    \right), \label{eq:sum_rate}
\end{align}
where $\gamma^{A}_{qnk}(\bm{B}) = \wa_{qnk}\herm\ch_{qnk}\herm\bm{B}^{-1}\ch_{qnk}\wa_{qnk}$ for $\bm{B}\in\C^{\nr\times\nr}$ is the signal-to-interference-noise ratio (SINR). Thus, we can now pose the following joint beamforming, scheduling and power allocation problem as
\begin{align}\tag{P1}\label{eq:general_problem}
    \max_{
    \substack{\alpha_{qnk}, p_{qnk}, \wa_{qnk} \\
    (n, k)\in\resourceset_{q}, q \in \userset
    }} &R^{A} \quad \text{s.t.} \\
    \tag{C1}\label{eq:binary}
    \alpha_{qnk} &\in \{0,1\} \\
    \tag{C2}\label{eq:pows_nonzero}
    \alpha_{qnk} p_{qnk} &\geq 0 \\
    \tag{C3}\label{eq:max_rbs}
    \sum_{(n,k)\in\resourceset_q}\alpha_{qnk} &\leq B_{q}\\ 
    \tag{C4}\label{eq:individual_powers}
    \sum_{(n,k)\in\resourceset_q}\alpha_{qnk}p_{qnk} &\leq P_{q}\\
    \tag{C5}\label{eq:precoders}
    \normsq{\wa_{qnk}} &\leq 1,
\end{align}
where $B_{q} = \lvert \resourceset_q \rvert$ is the maximum number of resource blocks which can be  allocated to user $q$ and $P_q$ the power constraints.
Note that the optimization problem \eqref{eq:general_problem} together with the constraints \eqref{eq:binary} to \eqref{eq:precoders} is a non-linear, non-convex, mixed-integer problem which is generally NP-hard due to the binary variables $\alpha_{qnk}$.
Furthermore, the metrics $R^{A}$ and $R^{B}$ depend on the noise covariances $\cov{\z_{nk}}$ which contain the jamming strategy and channel and are thus not known at the legitimate parties.

In our resilience framework, we assume the jammer has access to all system and physical parameters of all links and employs more transmit power and antennas than any legitimate party $q \in \mathcal{Q}$, i.e. $P_J \gg P_q$ and $N_J > N_{T_q}, N_{R}$.
The attacker thus seeks to find a jamming strategy for each RE $\cov{\bu_{nk}} \in N_{J}$ which minimizes the sum rate $R^{A}$. This is in contrast to prior works \cite{marti_mitigating_2023, marti_universal_2023}, which do not consider this worst-case scenario. More formally,
\begin{align}\tag{P2}\label{eq:prob_jammer}
    &\min_{\substack{ \cov{\bu_{nk}}  \in \C^{N_{J}\times N_{J}}
    \\ (n,k)\in\resourceset
    }} R^{A} \quad\text{s.t.}\\
    \tag{J1}\label{eq:jammer_spsd}
    &\cov{\bu_{nk}} = \cov{\bu_{nk}}\herm, \quad \cov{\bu_{nk}} \succeq \bm{0} \\ 
    \tag{J2}\label{eq:jammer_trace}
    &\sum_{(n,k) \in \resourceset}\tr{\cov{\bu_{nk}}}\leq P_{J},
\end{align}
where $\succeq$ denotes the L\"owner partial order.
The optimization problem above is a semidefinite program (SDP), which is a convex problem \cite{wolkowicz2012handbook}. 
Thus, a globally optimal solution can be found in polynomial time using interior-point or first-order methods. Both methods traditionally have the disadvantage that they do not scale well for high-dimensional SDPs like this one with a complexity of $O(N_{J}^4N^{2}K^{2})$ or higher \cite{wolkowicz2012handbook}.
Consequently, in Sections \ref{sec:legitimate_system_design} and \ref{sec:worst_case_jamming}, we will focus on answering following questions:
\begin{enumerate}
    \item How can the legitimate parties compute the filters in Equation \eqref{eq:user_rx} and solve the problem \eqref{eq:general_problem} efficiently in the absence of the true covariance matrix $\cov{\z_{nk}}$ using the sensing information $\bm{\theta}_{G}$?
    \item How can we find a scalable and efficient approximate solution to the worst-case jamming problem \eqref{eq:prob_jammer}?
\end{enumerate}

\section{Proposed Approach}\label{sec:proposed_approach}
\subsection{Resilient-by-Design Transceivers}\label{sec:legitimate_system_design}
In addition to the NP-hardness of \eqref{eq:general_problem}, the rate expression $R^A$ depend on the noise covariance matrix $\bm{C}_{\bm{z}_{nk}}$. Since the latter contains the jamming strategy, which is not known in the considered setup, a surrogate expression $\widetilde{\bm{C}}_{\bm{z}_{nk}}$ needs to be found, which effectively approximates the true covariance $\bm{C}_{\bm{z}_{nk}}$. To this end, it was shown in \cite{andrei_jcs_symp23} that such a surrogate covariance can be constructed using available information on the jamming signal DoAs, such that 
\begin{align}
    \widetilde{\bm{C}}_{\z_{nk}} = \eta \bm{A}(\bm{\theta}_G) \bm{A}(\bm{\theta}_G)^H + \sigma^2 \bm{I}_{N_R} \succeq \bm{C}_{\bm{z}_{nk}}, \label{eq:surrogate}
\end{align}
with the array manifold evaluated at the known DoAs, i.e.
\begin{align}
    \bm{A}(\bm{\theta}_G) = \begin{bmatrix} \bm{a}_{N_R}(\bm{\theta}_{G, 1}) \ \hdots \ \bm{a}_{N_R}(\bm{\theta}_{G, L_G}) \end{bmatrix}.
\end{align}
It is then straightforward to show that by inserting Eq. \eqref{eq:surrogate} into \eqref{eq:general_problem} and \eqref{eq:user_rx}, we maximize a lower bound on $R^{A}$ and $R^{B}$ respectively.
Now, instead of treating $\eta$ as unknown, we consider it a hyperparameter that controls the resilience level of our system. 
Since $\eta$ indicates the influence of the jamming signals, 
we conjecture that by choosing it much larger than the noise variance $\sigma^{2}$, we obtain approximately the same performance as if the true channel and jammer setup are known.  For the theoretical justification, we refer again to \cite{andrei_jcs_symp23}, and in Section \ref{sec:results} we will numerically demonstrate that this is the case for most scenarios considered. This approach is resilient by design since in the jammer-free case the parties adapt their system parameters with $\eta=0$ and if any illegitimate party is detected, $\eta$ is changed to a higher value. Thus the optimal receive filters in \eqref{eq:user_rx} can be computed and problem \eqref{eq:general_problem} can be solved using the surrogate covariance in \eqref{eq:surrogate}.

We now turn our attention to \eqref{eq:general_problem} and propose an iterative water-filling \cite{yu2004iterative} inspired algorithm for joint scheduling, power allocation, and beamforming shown in Algorithm \ref{alg:IWF_0}. Our main motivation for this is the fact that the original algorithm achieves capacity, as well as its excellent convergence properties. To our knowledge, there are no works extending this algorithm to account for scheduling. 
\begin{algorithm}
    \caption{Joint Iterative Scheduling, Beamforming, and Power Allocation}
    \label{alg:IWF_0}

    \textbf{Input:} Channels $\cg_{qnk}$, surrogate covariances $\widetilde{\bm{C}}_{\z_{nk}}$, power constraints $P_q$, maximum number of REs $B_q$

    \textbf{Output:} Wireless system design variables $\alpha_{qnk}$, $p_{qnk}$, $\bm{w}_{qnk}$

    \vspace{1mm} \hrule \vspace{1mm}

    \begin{algorithmic}[1]
        \STATE Initialize the interference-plus-noise covariance matrix to $\bm{X}_{qnk} = \widetilde{\bm{C}}_{\z_{nk}}$
        \STATE Initialize the design variables $\alpha_{qnk}^0$, $p_{qnk}^0$, $\bm{w}_{qnk}^0$ using the single user update in procedure in steps (6)-(11)
        
        \WHILE{not converged}
            \FOR{$q = 1$ to $Q$} 
                \STATE Update $\bm{X}_{qnk}$ using Equation \eqref{eq:interf_plus_noise} 
                \STATE Compute \textit{maximum eigenvalues} $\{\lambda_{qnk}\}_{(n,k) \in \mathcal{R}_q}$ of $\{\bm{H}_{qnk}^H \bm{X}_{qnk}^{-1} \bm{H}_{qnk}\}_{(n,k) \in \mathcal{R}_q}$
                and corresponding eigenvectors $\{\bm{t}_{qnk}\}_{(n,k) \in \mathcal{R}_q}$
                \STATE Determine indices $\mathcal{I}_q$ of the largest $B_q$ eigenvalues
                \STATE Set $\alpha_{qnk} = 1$ for all $(n,k) \in \mathcal{I}_q$ and 0 else
                \STATE Compute $p_{qnk} = (\mu - \lambda_{qnk}^{-1})^{+}$ with $\mu$ chosen such that $\sum_{\mathcal{I}_q} p_{qnk} \leq P_q$
                \STATE Set $\bm{w}_{qnk} = \bm{t}_{qnk}$ for $(n,k) \in \mathcal{I}_q$
            \ENDFOR
        \ENDWHILE
    \end{algorithmic}
\end{algorithm}
In our case, each user computes the matrices $\bm{X}_{qnk}$ and updates its variables according to
\begin{align}\nonumber
    &\max_{
    \substack{\alpha_{qnk}, p_{qnk}, \wa_{qnk} \\
    (n, k)\in\resourceset_{q}
    }} \sum_{\substack{(n,k)\in\resourceset}} 
    \alpha_{qnk}\log\left( 1 + p_{qnk} \gamma_{qnk}^{A}(\bm{X}_{qnk}) \right) \\
    & \text{s.t. \eqref{eq:binary} to \eqref{eq:precoders}} \label{eq:iwf_step}
\end{align}
using lines 6 to 11 in Algorithm \ref{alg:IWF_0} until convergence, with $\gamma_{qnk}^{A}(\bm{X}_{qnk})$ as in Sec. \ref{sec:legitimate_system_design}.
In order to see why the proposed method will offer good performance, first note that the sum rate is maximized if we maximize the sum rate of all resource elements individually. 
Secondly, as already stated, the iterative water-filling algorithm \cite{yu2004iterative} in its' original formulation converges quickly to the sum capacity of the MIMO MAC, and thirdly it can be easily shown that the proposed single-user updates given by lines 6 to 11 in Algorithm \ref{alg:IWF_0} find the globally optimal solution to the problem in \eqref{eq:iwf_step}.
Note that the complexity of this algorithm is dominated by the computation of the maximum eigenvalue and eigenvector. This can be solved by a power iteration algorithm which scales quadratically in the number of variables. Thus the total complexity scales as $O(N_{\text{iter}} Q \nt^{2} N K)$ in both time and memory, but note that the innermost loop can be paralellized over the subcarriers and OFDM symbols. 
Finally, a preliminary experimental convergence analysis shows that the proposed algorithm reaches convergence after very few iterations. Due to lack of space, we defer it to future work.  
\subsection{Worst-Case Jamming Strategy}\label{sec:worst_case_jamming}
In order to benchmark the proposed sensing-assisted resilience framework even against the most disruptive kind of adversarial jammers, we now address the question of finding a computationally efficient approximate solution to the problem 
\eqref{eq:prob_jammer}. To this end, a two-step approximation procedure is proposed in Algorithm \ref{alg:WCJ}, consisting of a prior user selection stage and a closed form computation of the jamming convariances, which is derived in the following.

%

\begin{algorithm}
    \caption{Approximate Worst-Case Jamming Strategy}
    \label{alg:WCJ}

    \textbf{Input:} Legitimate channel $\bm{H}_{qnk}$, jamming channel $\bm{G}_{nk}$, jamming power budget $P_J$ 
    
    \textbf{Output:} Worst-case jamming covariance matrix $\bm{C}_{\bm{u}_{nk}}$
    \vspace{1mm} \hrule \vspace{1mm}

    \begin{algorithmic}[1]
        \STATE For each user $q \in \mathcal{Q}$, compute the alignment matrix \\ $\bm{R}_{qnk} = \bm{G}_{nk}^{\dagger} \bm{H}_{qnk} \bm{H}_{qnk}^H \bm{G}_{nk}^{\dagger, H}$ 
        
        \STATE Determine the strongest user \\ $q^* = \arg \max_{q \in \mathcal{Q}} \lambda_{\max}(\bm{R}_{qnk})$

        \STATE For the strongest user alignment $\bm{R}_{q^*nk}$, compute the eigenvectors $\bm{U}_{q^*nk}$ and eigenvalues $\{\lambda_{q^*nk,d}\}_{d=1}^{N_J}$

        \STATE Compute the jamming power allocation weighting \\ $h_{nk} = p_{q*nk} \sum_{d=1}^{N_J} \lambda_{q*nk,d}$

        \STATE Compute the jamming power scaling factors \\ $g_{nk} = \left ( P_J \sqrt{h_{nk}} \right ) / \left ( \sum_{nk} \sqrt{h_{nk}} \right )$

        \STATE Compute the jamming power allocation matrix \\ $\bm{\Lambda}_{\bm{u}_{nk}} = \text{diag}\left\{ \dfrac{g_{nk} \lambda_{q^*nk, d}}{\sum_{d=1}^{N_{J}} \lambda_{q^*nk, d}}  \right\}_{d=1}^{N_{J}}$ 

        \STATE Compute the jamming covariance matrix \\ $\bm{C}_{\bm{u}_{nk}} = \bm{U}_{q^*nk} \bm{\Lambda}_{\bm{u}_{nk}} \bm{U}_{q^*nk}^H$

    \end{algorithmic}
\end{algorithm}

The corresponding user selection stage is motivated by the importance of the strongest users in terms of SINR $\gamma^{A}_{qnk}$ for each RE. 
Using standard linear algebra arguments and Jensen's inequality one can show that
\begin{equation} \label{eq:jamm_upper_bound}
        R^{A} \leq \lvert\resourceset_{J}\rvert\log\left(1 + \frac{Q}{ \lvert\resourceset_{J}\rvert} \sum_{(n,k) \in \resourceset_{J}} \gamma_{\text{max}}\left(\cov{\z_{nk}}\right) \right)
    \end{equation}
where $
        \gamma_{\text{max}}\left(\cov{\z_{nk}}\right) = \max_{q\in\userset} p_{qnk}\lambda_{\text{max}}\left(\ch_{qnk}\herm\cov{\z_{nk}}^{-1}\ch_{qnk}\right)
$
and $\mathcal{R}_{J}$ is the set of scheduled REs.
Accordingly, disrupting the latter for each RE implies a deterioration of the total sum-rate. Furthermore, as the worst-case jammer is assumed to be in the jammer-dominant regime, it follows from $P_J \gg \sigma^2$ that
\begin{align}
    \bm{C}_{\bm{z}_{nk}} = \bm{G}_{nk} \bm{C_{\bm{u}_{nk}}} \herm{\bm{G}_{nk}} + \sigma^2 \bm{I}_{N_R} \approx \bm{G}_{nk} \bm{C_{\bm{u}_{nk}}} \herm{\bm{G}_{nk}}.
\end{align}
We now introduce the alignment matrix
\begin{align}
    \bm{R}_{qnk} = \bm{G}_{nk}^{\dagger} \bm{H}_{qnk} \bm{H}_{qnk}^H \bm{G}_{nk}^{\dagger, H},
\end{align}
where $(\cdot)^{\dagger}$ denotes the Moore-Penrose pseuodinverse, to quantify impact of jamming on the corresponding REs.
Using the analysis in \cite[Proposition 1]{disrupting_mimo_comms} we have that
\begin{align}
    \lambda_{\max} \left ( \bm{H}_{qnk}^H \bm{C}_{\bm{z}_{nk}}^{-1} \bm{H}_{qnk} \right ) \approx \lambda_{\max} \left ( \bm{R}_{qnk} \bm{C}_{\bm{u}_{nk}}^{-1} \right ),
    \label{eq:lambda_approximation}
\end{align}
with $\lambda_{\max}(\mathbf{A})$ the largest eigenvalue of the matrix $\mathbf{A}$. 
Accordingly, the jammer will choose the strongest user $q^*$ as 
\begin{align}
    q^* = \arg \max_{q \in \mathcal{Q}} p_{qnk} \lambda_{\text{max}} (\bm{R}_{qnk}).
    \label{eq:user_selection_stage}
\end{align}

With this user selection stage, the original optimization problem in \eqref{eq:prob_jammer} can be recast as  
\begin{alignat}{2}
    \min_{
    \substack{\bm{C}_{\bm{u}_{nk}} }
    } \quad \lambda_{\max} \left ( \bm{R}_{q^*nk} \bm{C}_{\bm{u}_{nk}}^{-1} \right ) \quad &\text{s.t.} \quad \eqref{eq:jammer_spsd} \tag{P3} \label{eq:P3} \\
    g_{nk} &\geq 0 \tag{K1} \label{eq:K1} \\
    \trace ( \bm{C}_{\bm{u}_{nk}} ) &\leq g_{nk}, \tag{K2} \label{eq:K2}
\end{alignat}
where the constraint \eqref{eq:K1} refers to the non-negativeness of the power scaling factors $g_{nk}$ and \eqref{eq:K2} incorporates these into a power allocation constraint for each resource element. Due to the previous user selection stage, user $q^*$ and its resource elements $(n,k)$ are iteratively fixed. A corresponding closed-form solution to the alternative problem statement in \eqref{eq:P3} has been derived in \cite[Proposition 2]{disrupting_mimo_comms} as
\begin{align}
    \bm{C}_{\bm{u}_{nk}} &= \bm{U}_{q^*nk} \bm{\Lambda}_{\bm{u}_{nk}} \bm{U}_{q^*nk}^H.\label{eq:WC_solution}
\end{align}

In this expression, $\bm{U}_{q^*nk}$ denotes the eigenvector matrix of the alignment $\bm{R}_{q^*nk}$ and $\bm{\Lambda}_{\bm{u}_{nk}}$ denotes the jamming power allocation matrix with $\lambda_{q^*nk,d}$ being the $d$-th eigenvalue of $\bm{R}_{q^*nk}$, i.e. 
\begin{align}
    \bm{\Lambda}_{\bm{u}_{nk}} &= \text{diag} \left \{ \frac{g_{nk} \lambda_{q^*nk,d}}{\sum_{d=1}^{N_J} \lambda_{q^*nk,d}} \right \}_{d=1}^{N_J} \label{eq:big_lambda}
\end{align}

The remaining power scaling factors $g_{nk}$ are determined by solving the convex subproblem
\begin{align}
    \min_{
    \substack{g_{nk} \\ (n,k) \in \mathcal{R}_J}
    } \quad \sum_{(n,k) \in \mathcal{R}_J} \frac{h_{nk}}{g_{nk}} \quad &\text{s.t.} \quad \eqref{eq:K1} \tag{P4} \label{eq:P4} \\
    \sum_{(n,k) \in \mathcal{R}_J} g_{nk} &\leq P_J, \tag{L1} \label{eq:L1}
\end{align}
with the jamming power allocation $h_{nk} = p_{q^*nk} \sum_{d=1}^{N_J} \lambda_{q^*nk,d}$. This can be analytically solved using Lagrange multipliers with the solution being given in lines 4 and 5 of Algorithm \ref{alg:WCJ}. Note that the complexity of this algorithm is dominated by the eigenvalue decomposition, which scales as $O(N_{J}^{3})$ per RE. 

Taking into account these factors, the problem originally presented in \eqref{eq:P3}, which is convex but scales as $O(N_{J}^4N^{2}K^{2})$, has been converted into an approximate problem which scales as $O(N_{J}^3 NK)$. This latter is addressed through a sequence of closed-form solutions, following an initial user selection stage. In Section \ref{sec:results}, we will see that this proposed strategy effectively nullifies user-sum rates in all unprotected cases.

\section{Simulation Results and Analysis} \label{sec:results}
\subsection{Wireless Setup and Methodology}
\begin{figure}
    \includegraphics[scale=0.4]{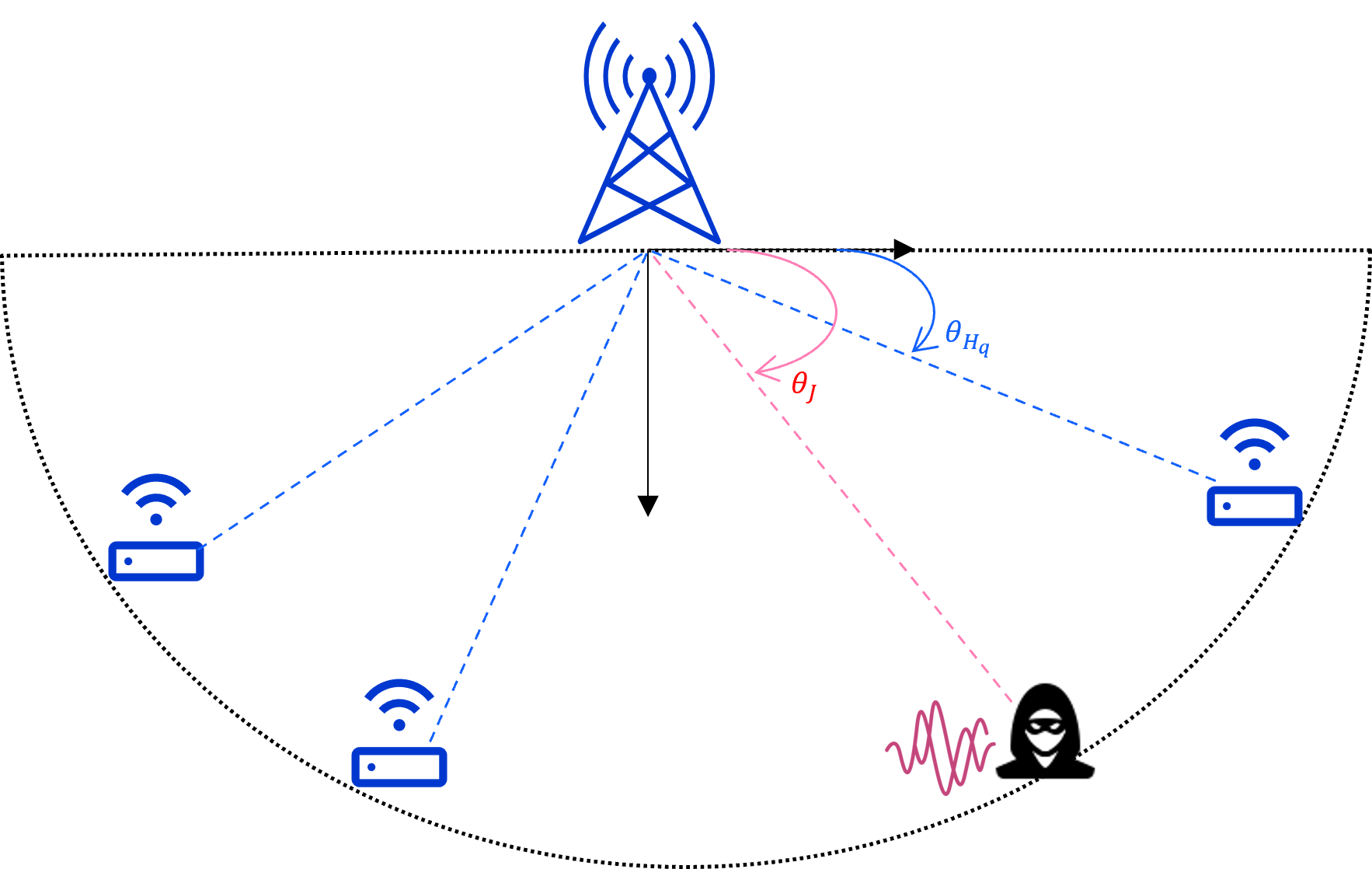}
    \centering
    \caption{Qualitative rendering of the considered scenario.}
    \label{fig:setup}
\end{figure}

Consider the setup depicted in Fig. \ref{fig:setup}, where both the UEs and the jammer are positioned on a circle in the $xy$-plane, with the BS at the center. The azimuth DoAs for the UEs are set up with a central angle $\theta_{H_{q}}$ spaced by $5^{\circ}$ and an additional random disturbance $\phi_l$ drawn from a uniform distribution $\mathcal{U}(-5^\circ, 5^\circ)$. A similar method is applied for the jamming channel. We consider uniform linear arrays with $\ntq = 8$ for all transmitters and $\nr = 16$ for the receiver. We model our slots similar to LTE, with $K=14$ symbols and $N=64$ subcarriers. The number of resource blocks per user $B_{q}$ is set to $\lfloor\frac{NK}{Q}\rfloor$, and each user has a power budget of $P_{q} = 5\si{\dBm}$. For this simulation, the number of users is $Q=3$, the design hyperparameter $\eta = 10$, and the noise level $\sigma^{2}=-3\si{\dBm}/\si{Hz}$, with all channels computed using equations \eqref{eq:beamspace1} and \eqref{eq:beamspace2}.

In the considered setup, the UEs calculate the surrogate noise covariance matrix $\widetilde{\bm{C}}_{\z_{nk}}$ using equation \eqref{eq:surrogate}. They then optimize their system parameters, which are subsequently utilized to determine the worst-case jamming strategy and to compute the user sum-rate $R^{B}$. Contrary to approaches as in \cite{marti_mitigating_2023, marti_universal_2023}, our methodology assumes that while the jammer is aware of the optimized system parameters, the legitimate parties only have knowledge of the jamming signal's DoAs. Three baseline scenarios are considered for comparison:
\begin{enumerate}
    \item \textit{No Jammer}: Evaluating the system in the absence of any jamming activity.
    \item \textit{No Protection}: The jammer is active, but no specific protection is employed against it.
    \item \textit{Full Knowledge}: The system has complete knowledge of the jammer's channel and strategy, and optimizations are performed accordingly.
\end{enumerate}
To showcase the effectiveness of our proposed jamming strategy, results also include scenarios with a barrage jammer, where $\cov{\bu_{nk}} = P_{J} / (N_{J}N K)$. 
\begin{figure*}
    \begin{subfigure}{\columnwidth}
        \centering
        \includegraphics[width=\columnwidth]{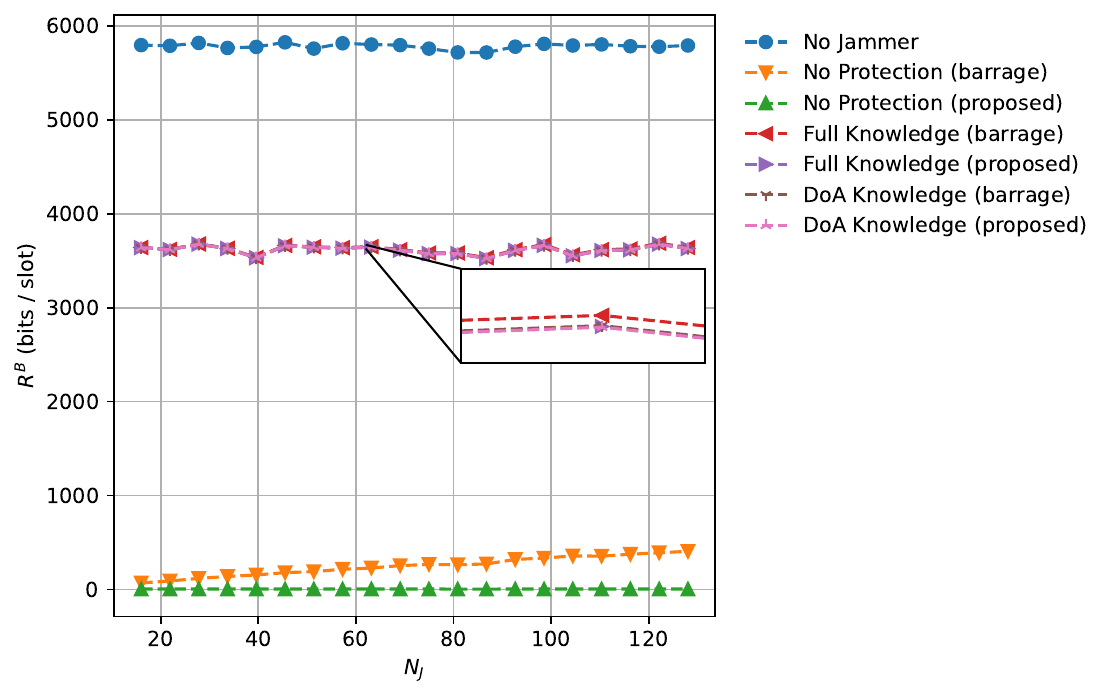}
        \caption{User sum-rate vs. number of antennas.
        $P_{J} = 30\si{\dBm}$, $\theta_J = 20^{\circ}$.}
        \label{fig:antennas}
   \end{subfigure}
   \hfill   
    \begin{subfigure}{\columnwidth}
        \centering
        \includegraphics[width=\columnwidth]{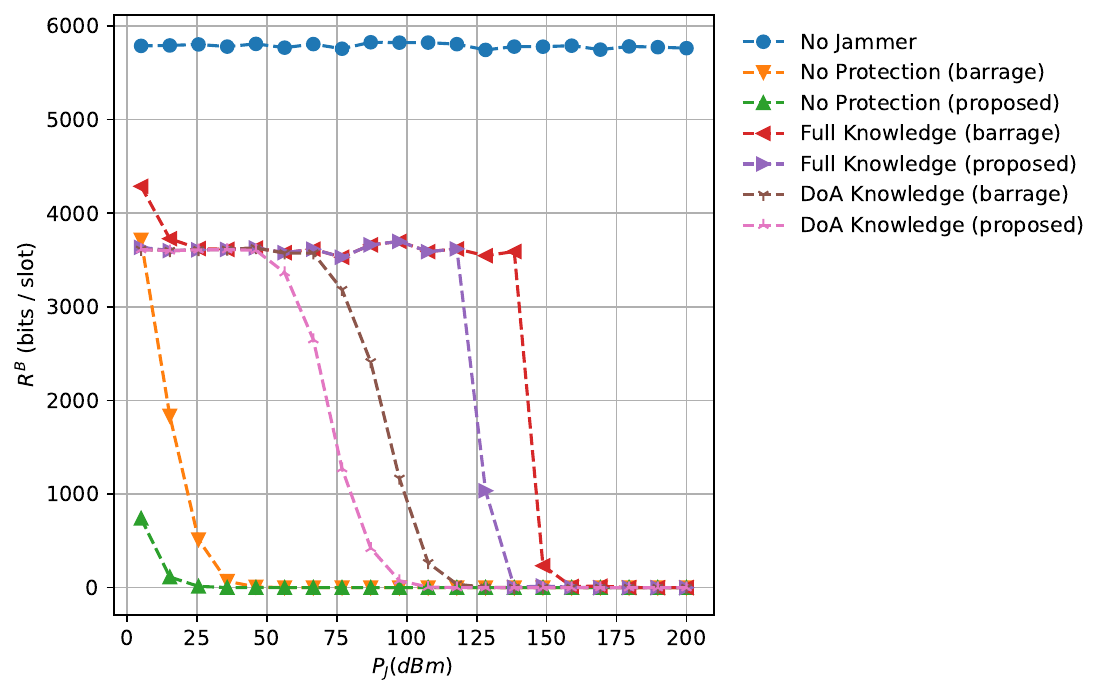}
        \caption{User sum-rate vs. jammer powers, $N_{J} = 64$, $\theta_{J} = 20^{\circ}$.}
        \label{fig:powers}
    \end{subfigure}
    \vfill
    \begin{subfigure}{\columnwidth}
        \centering
        \includegraphics[width=\columnwidth]{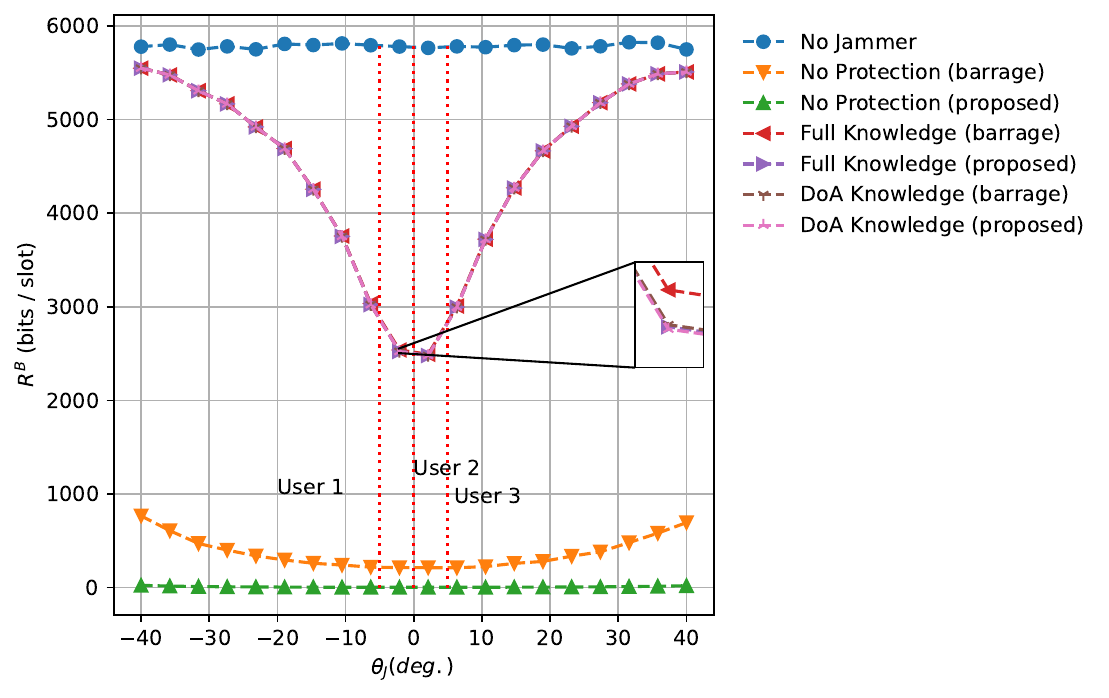}
        \caption{User sum-rate vs. jamming DoAs, 
        $N_{J} = 64$, $P_{J} = 30\si{\dBm}$.}
        \label{fig:aoas}
    \end{subfigure}
    \hfill
    \begin{subfigure}{\columnwidth}
        \centering
        \includegraphics[width=\columnwidth]{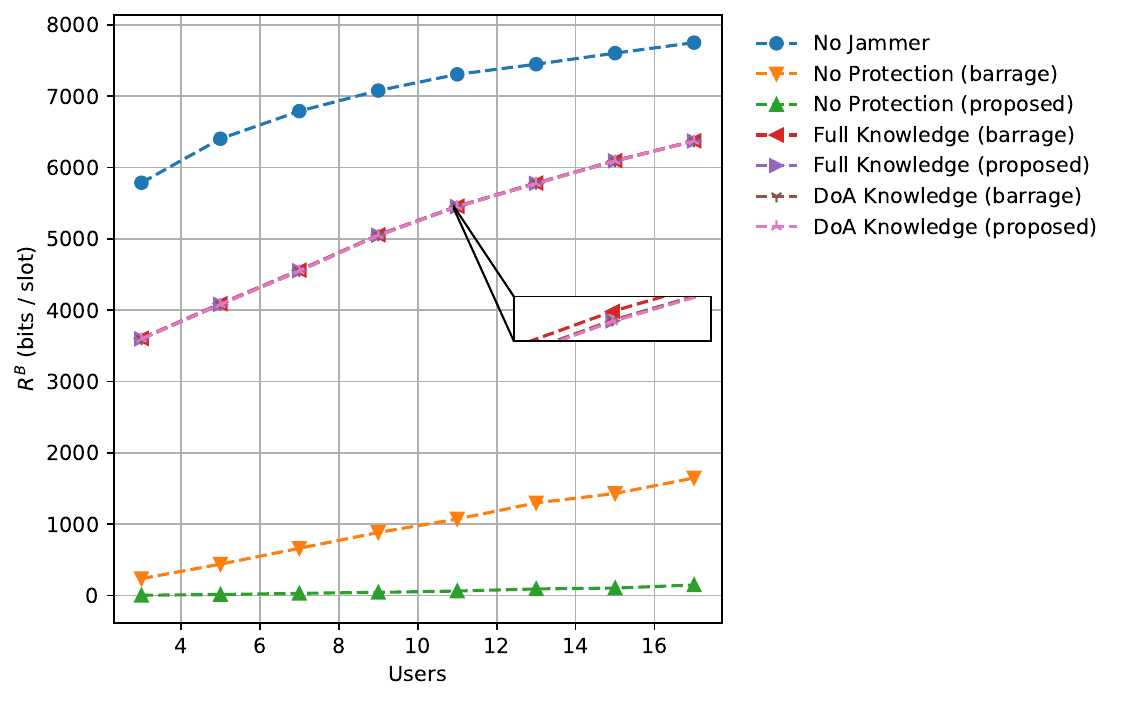}
        \caption{User sum-rate vs. number of users, 
        $N_{J} = 64$, $P_{J} = 30\si{\dBm}$.}
        \label{fig:users}
    \end{subfigure}
    \caption{Simulation results for all considered scenarios.}
    \label{fig:results}
\end{figure*}

\subsection{Impact of Number of Antennas}
The study first examines how the number of jammer antennas affects system resilience. As shown in Figure \ref{fig:antennas}, with protective measures in place, the sum user rate remains consistent, unaffected by the number of jammer antennas. This stability represents a significant rate improvement compared to worst-case scenarios. Moreover, optimizing transmit and receive parameters based on DoA information matches performance levels akin to scenarios with full knowledge of the jammer's setup. The proposed jamming strategy also demonstrates greater effectiveness than barrage jamming in unprotected scenarios, keeping user rates nearly zero across a wide range of antenna numbers.
\subsection{Impact of Jamming Power}
When varying the jamming power, our methods ensure resilience even at extremely high jamming power levels. However, a threshold exists beyond which the knowledge of jammer DoAs alone is insufficient, leading to a notable decline in user sum rates, observable in Fig. \ref{fig:powers}. This decline happens at higher jamming powers when full jammer knowledge is available. In these scenarios, the proposed jamming strategy remains more effective than standard barrage jamming, effectively reducing communication rates to near zero with even modest jamming power budgets.
\subsection{Impact of Jamming DoAs}
Exploring the dependency on adversarial signal DoAs involves analyzing three legitimate users with varying central DoAs, as depicted in \ref{fig:aoas}. User rates tend to decrease as jamming signals align more closely in direction. Yet, even with closely aligned jamming signals, the system maintains a transmission rate of about $2500$ bits per slot in protected scenarios. Performance with full jammer knowledge parallels that achieved using only DoA knowledge. The proposed jamming strategy continues to stand out, reducing user rates to zero across all DoAs, demonstrating its efficacy in approximating the original semidefinite program solution.
\subsection{Impact of Number of Users}
The final aspect of the study considers the effect of an increasing number of users. Results in Figure \ref{fig:users} confirm the advantages of our anti-jamming approaches, consistently outperforming unprotected scenarios. Notably, the proposed jamming strategy effectively nullifies user rates, regardless of the user count, underscoring its robustness against varying system scales.

\section{Conclusions} \label{sec:conclusions}
In this work, we have introduced a sensing-assisted and resilient-by-design approach for anti-jamming in MIMO-OFDM-MAC systems and have incorporated this methodology into the joint optimization of beamforming, scheduling, and power allocation. Our proposed method operates effectively and solely with information on the adversary's DoAs and significantly enhances system performance in various jamming scenarios, consistently outperforming unprotected setups under diverse channel conditions and jamming configurations. This holds even for the worst-case jammer, for which a computationally efficient approximation of the optimal startegy was derived. Future work will focus on the optimality and convergence properties of our resource allocation scheme, on a comparative analysis between our approximate jamming strategy and its globally optimal but not scalable counterpart, as well as on the practical validation of our approaches in field tests. 
\bibliography{IEEEabrv, references}
\bibliographystyle{IEEEtran}
\end{document}